# The New SARS-CoV-2 Strain Shows a Stronger Binding Affinity to ACE2 Due to N501Y Mutation

Fedaa Ali[1], Amal Kasry[1] and Muhamed Amin[2,3] *

[1] Nanotechnology Research Centre (NTRC), the British University in Egypt (BUE), El Sherouk City, Suez Desert Road, Cairo 1183.
[2] Department of Sciences, University College Groningen, University of Groningen, Hoendiepskade 23/24, 9718 BG Groningen, The Netherlands.
[3] Center for Free-Electron Laser Science, Deutsches Elektronen-Synchrotron DESY, Notkestrasse 85, 22607 Hamburg, Germany.

* Corresponding author: m.a.a.amin@rug.nl

**ABSTRACT:** SARS-CoV-2 is a global challenge due to its ability to spread much faster than SARS-CoV, which was attributed to the mutations in the receptor binding domain (RBD). These mutations enhanced the electrostatic interactions. Recently, a new strain was reported in the UK that includes a mutation (N501Y) in the RBD, that possibly increases the infection rate. Using Molecular Dynamics simulations (MD) and Monte Carlo (MC) sampling, we showed that the N501 mutation enhances the electrostatic interactions due to the formation of a strong hydrogen bond between SARS-CoV-2-T500 and ACE2-D355 near the mutation site. In addition, we observed that the electrostatic interactions between the SARS-CoV-2 and ACE2 in the wild type and the mutant are dominated by salt-bridges formed between SARS-CoV-2-K417 and ACE2-D30, SARS-CoV-2-K458, ACE2-E23, and SARS-CoV-2-R403 and ACE2-E37. These interactions contributed more than 40 % of the total binding energies.

SARS-CoV-2 is the infectious agent of the highly spreading coronavirus disease 2019. Since the beginning of 2020, the number of infections is spreading numerously in almost everywhere around the world[1–4]. Currently, the total number of confirmed cases reached more than 80 million cases and nearly 2 million deaths in more than 190 countries.

Previous studies reported that the angiotensin converting enzyme 2 (ACE2) is the receptor, which facilitates its binding and entry to the host cells[5,6]. Recently, it was shown that the spread rate of the virus has become much faster due to different genetic changes in the receptor-binding domain and the FURIN cleavage site [7]. These changes include the mutation of Asparagine at position 501 to Tyrosine (N501Y), which is one of the residues in the RBD-ACE2 contact area. Experimental findings showed that the N501Y could enhance the binding affinity of Sars-Cov2 spike protein to ACE2.[8,9]

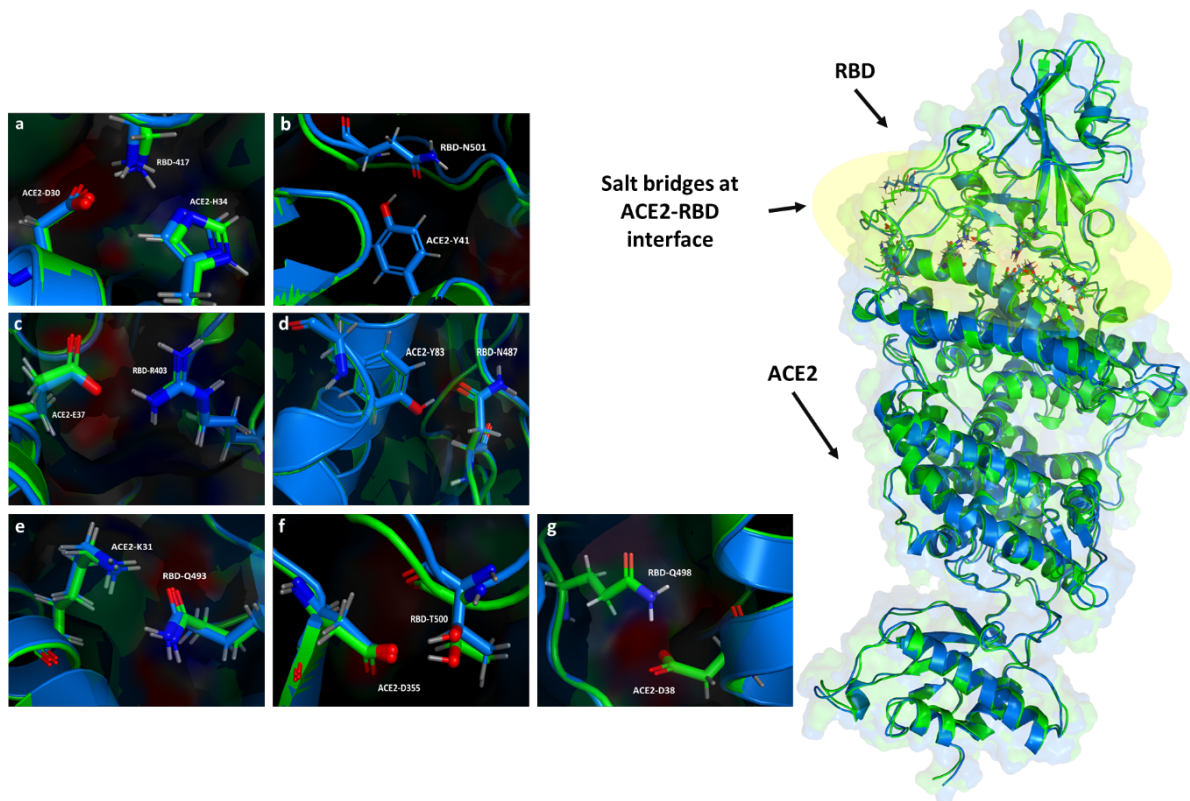

**Figure 1.** Salt-bridges at between RBD and ACE2 in both of WT and N501Y mutated structure. The WT RBD and ACE2 are shown in Blue while the N501Y mutated RBD and ACE2 are shown in Green. (a-g) Different salt-bridges interactions (interaction energies are shown in Table.2)

Binding affinity of viruses to the host receptors is mainly affected by different protein – protein electrostatic interactions[10,11]. A change in different pivotal residues at the binding site could affect viral-host cells fusion and hence the infectivity of the virus[5,12–14]. Therefore, we herein study how the N501Y mutation in the RBD of the SARS-CoV2 can alter the binding of the virus to ACE2. We focus on the role of the

**Table 1.** The interaction energies between SARS-CoV-2-RBD and ACE2 in both WT and N501Y mutated structures.

|       | Coulomb (kcal/mol) | Van der Waals (kcal/mol) | Total (kcal/mol) |
|-------|--------------------|---------------------------|------------------|
| Wt    | -18.38             | -31.56                    | -49.94           |
| N501Y | -21.91             | -32.59                    | -53.91           |

The electrostatic interactions are calculated for the optimized most occupied conformer of the proteins by solving Poisson Boltzmann equation.

electrostatic interactions on the binding energy, where it is known to be dominant among different protein-protein interactions[15]. In this study, we used a combined Molecular dynamic (MD) and Monte Carlo (MC) simulations to assess the molecular interactions between RBD of S-protein and ACE2 for the N501Y mutant and compared our results with the wild type. The crystal structure (PDB ID: 6M17)[12] was optimized using openMM[16]. Then, several rotamers were built by MCCE to by rotating each rotatable bond by 60º to appropriately sample the sidechains conformations. In order to build the N501Y mutant, the sidechain of N501 is replaced by aromatic sidechain of tyrosine using MCCE.

The electrostatic interactions between the different conformers were calculated using DELPHI.[17] Then, MCCE is used to generate Boltzmann distribution for all conformer using MC sampling for the wild type and the N501Y mutant at pH 7. The most occupied conformers were subjected to MD minimization again using openMM.[16] The resulting structures were used to calculate the electrostatic energies between the RBD of SARS-CoV-2 and the ACE2 using DELPHI. The electrostatic interactions between the SARS-CoV-2 and ACE2 in the wild type and the mutant are dominated by salt-bridges formed between SARS-CoV-2-K417 and ACE2-D30, SARS-CoV-2-K458, ACE2-E23, and SARS-CoV-2-R403 and ACE2-E37 (Figure 1). These residues contribute by ~20 Kcal/mol of the total electrostatic interactions between the SARS-CoV-2 and ACE2 (Table S1). However, the binding between the SAR-CoV-2 and ACE2 is favored by ~4 Kcal/mol in the mutant (Table 1) due to a stronger hydrogen bond between SARS-CoV-2-T500 and ACE2-D355 (Figure 2a, Table S1).

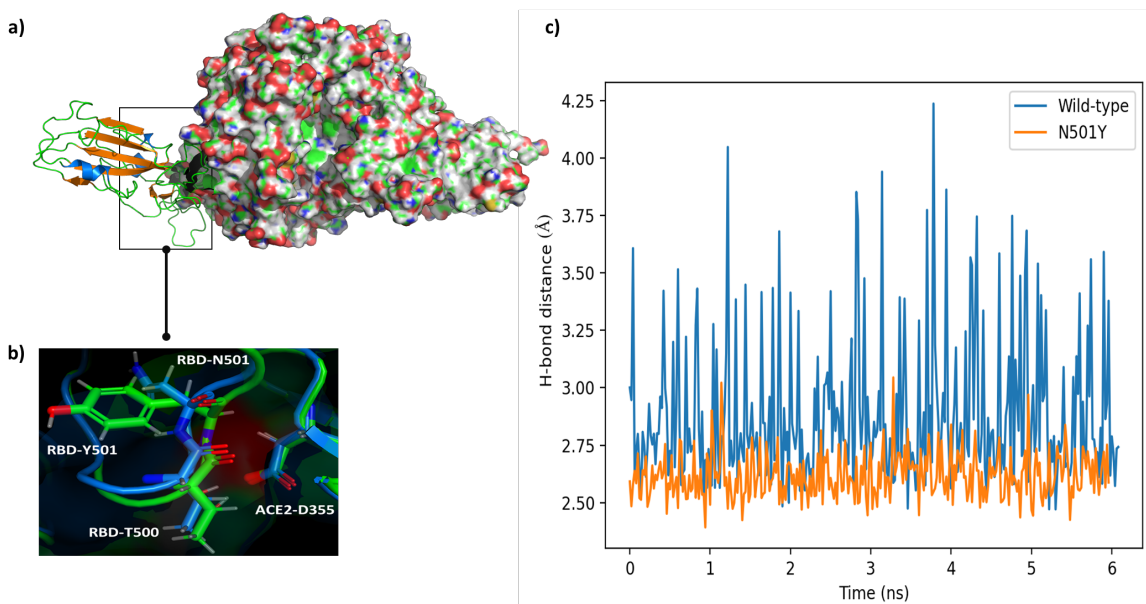

**Figure 2. (a)** [PDB code 6M17]: SARS-CoV2 RBD and human ACE2 complex. RBD is shown in cartoon view. **(b)** Residues at position 501 for both of WT and N501Y mutated structures and interacting residue in the vicinity of position 501.The WT RBD and ACE2 are shown in Blue while the N501Y mutated RBD and ACE2 are shown in Green. **(c)** The distance in Å between OG1 of SARS-CoV-2 and ACE2-T500 and OD2 of D355 through 6ns MD trajectory.

In WT, the maximum vdW interaction of -1.889 kcal/mol was reported between residue RBD-N501 and ACE2-K353. While For the mutated structure, a maximum vdW interactions of -2.441 Kcal/mol was between RBD-Q498 and ACE2-Y41.

The Maximum electrostatic interactions of -9.55 kcal/mol was observed between RBD-K417 and ACE2-D30 in wild type, and of -8.39 kcal/mol between RBD-K458 and ACE2-E23 in mutated structure. The N501Y mutant is shown to decrease the repulsion between carboxylate of ACE2-D355 and backbone oxygen of RBD-T500 RBD-T500 as the distance decreased from 3.81 to 3.42 Å.

In addition, the hydrogen bond between the carboxylate of ACE2-D355 and HG1 of RBD-T500 is shortened from 1.65 to 1.57Å in the mutant. To further asses the stability of this hydrogen bond, we ran MD trajectories for 6 ns Figure. 1(c). The trajectories show that there is a significant variation in the bond length in the wild type, while in the mutant, the variation is much less due to the strong interactions. However, the average distance in between OG1 of SARS-CoV-2 and ACE2-T500 and OD2 of D355 is shorter for the mutant (2.61Å) compared to the wild type (2.85Å).

In summary, we showed that the binding affinity of SARS-CoV-2 to human ACE2 is larger in the N501Y mutated structure than that in WT because of the significant change in the electrostatic interactions. Upon mutation, the salt-bridge electrostatic interaction increased between T500 and D355 in the RBD and ACE2, respectively, to be ~3.39 kcal/mol more negative than that in the WT. This is shown to be due to the shorter atomic distances between ACE2-D355 and SARS-CoV2-T500.

**Acknowledgement**
This research was supported in part through the Maxwell computational resources operated at Deutsches Elektronen-Synchrotron DESY, Hamburg, Germany. We would like to thank Prof. Jochen Kupper for useful discussion.

# Supporting Information for:

# The New SARS-CoV-2 Strain Shows a Stronger Binding Affinity to ACE2 Due to N501Y Mutation

Fedaa Ali[1], Amal Kasry[1], Muhamed Amin[2,3] *

[1] Nanotechnology Research Centre (NTRC), the British University in Egypt (BUE), El Sherouk City, Suez Desert Road, Cairo 1183.
[2] Department of Sciences, University College Groningen, University of Groningen, Hoendiepskade 23/24, 9718 BG Groningen, The Netherlands.
[3] Center for Free-Electron Laser Science, Deutsches Elektronen-Synchrotron DESY, Notkestrasse 85, 22607 Hamburg, Germany.

* Corresponding author: m.a.a.amin@rug.nl

The supplementary Information index:

**S1:** Details of the MD and MC simulations

**S2:** Contribution of electrostatic and vdW interactions between RBD and ACE2 in both WT and N501Y mutated structures

**S3:** The PDB files of the optimized structures of the wild type and the N501Y mutant is attached as a separate file.

### S1: Computational Methods

The crystal structural is obtained from protein data bank (PDB: 6M17[1]) then it was optimized with openMM[2] using Amber99sb forcefield and implicit water model for the solvent. Starting from the optimized structure, MCCE[3] is used to generate protonation conformers for the charged amino acids in the structure. Poisson-Boltzmann electrostatic interactions and the solvation energies are calculated using Delphi[4], while the Lennard–Jones energies are obtained from the AMBER forcefield.[5] The protein interior is assigned a dielectric constant of 4 and the probe radius used for calculating solvent accessible surface is 1.4 Å. The ion probe radius is 2.00 Å and the salt concentration is 0.15 mM.

Monte Carlo sampling is used to generate an equilibrium ensemble of conformational microstates for conformers with different protonation patterns at pH 7 according to the following free energy ΔG.

$$\Delta G^x = \sum_{i=1}^{M} \delta_{x,i}\{[2.3 m_i K_b T(7.0 - pK_{a,sol,i})] + \Delta\Delta G_{solv,i} + \sum_{i=1+1}^{M} \delta_{x,i}[\Delta G_{ij}]\}$$

M is the total number of conformers, $\delta_{x,i}$ is the fragment state, where it equals 1 if the conformer i is existing and 0 if not. $K_b T$ 25.37 meV at 298 K, $n_i$ is the number of electrons gained, F is the faraday constant. $pK_{a,sol,i}$ and $E_{m,sol,i}$ are the pKa and the reduction potential $E_m$ for each fragment "i" in the reference dielectric medium. $\Delta\Delta G_{solv,i}$ is the resultant free energy for the movement of a fragment from water into a cluster in the same solvent. $\Delta G_{ij}$ is the free energy coming from both the electrostatic and the Lennard-Jones interaction of fragments "i" and "j" in the cluster of a microstate x.

**S2: Table S1.** Contribution of electrostatic and vdW interactions between RBD and ACE2 in both WT and N501Y mutated structures

| Wild Type RBD-ACE2 complex | | | | | | N501Y mutated RBD-ACE2 complex | | | | | |
|---|---|---|---|---|---|---|---|---|---|---|---|
| Electrostatic energy | | | LJ energy | | | Electrostatic energy | | | LJ energy | | |
| RBD | ACE2 | Energy | RBD | ACE2 | Energy | RBD | ACE2 | Energy | RBD | ACE2 | Energy |
| K 417 | D 30 | -9.55 | N 501 | K 353 | -1.889 | K 458 | E 23 | -8.39 | Q 498 | Y 41 | -2.441 |
| K 458 | E 23 | -8.08 | Q 498 | Y 41 | -1.785 | K 417 | D 30 | -7.98 | F 456 | T 27 | -1.392 |
| R 403 | E 37 | -3.96 | N 487 | Q 24 | -1.468 | <span style="color:red">T 500</span> | <span style="color:red">D 355</span> | <span style="color:red">-5.51</span> | N 487 | Q 24 | -1.253 |
| N 501 | Y 41 | -2.15 | F 456 | T 27 | -1.415 | R 403 | E 37 | -4.14 | L 455 | K 31 | -1.228 |
| N 487 | Y 83 | -2.13 | L 455 | K 31 | -1.169 | Q 498 | D 38 | -3.49 | Y 453 | H 34 | -1.203 |
| <span style="color:red">T 500</span> | <span style="color:red">D 355</span> | <span style="color:red">-2.12</span> | Q 498 | L 45 | -1.13 | N 487 | Y 83 | -2.18 | Y 501 | Y 41 | -1.036 |
| Q 493 | K 31 | -1.95 | Y 489 | K 31 | -1.121 | Q 493 | K 31 | -1.93 | F 456 | D 30 | -1.028 |
| K 417 | H 34 | -1.57 | Y 453 | H 34 | -1.084 | K 417 | H 34 | -1.53 | Y 505 | K 353 | -1.015 |
| R 403 | D 38 | -1.11 | L 455 | H 34 | -1.075 | K 417 | E 37 | -1.02 | L 455 | H 34 | -0.968 |
| D 405 | K 353 | -0.98 | T 500 | R 357 | -0.982 | D 405 | K 353 | -0.94 | P 499 | L 45 | -0.94 |
| E 406 | D 38 | 0.52 | T 500 | N 330 | 0.136 | K 417 | R 393 | 0.56 | Q 498 | D 38 | 0.089 |
| N 501 | D 355 | 0.52 | T 500 | D 355 | 0.17 | D 405 | D 30 | 0.56 | Y 489 | T 27 | 0.165 |
| D 405 | D 30 | 0.55 | N 487 | Y 83 | 0.271 | K 417 | K 353 | 0.63 | K 417 | H 34 | 0.379 |
| K 417 | K 353 | 0.63 | Q 493 | K 31 | 0.528 | E 406 | D 30 | 0.8 | N 487 | Y 83 | 0.391 |
| N 501 | K 353 | 0.74 | Q 493 | E 35 | 0.871 | R 403 | R 393 | 0.92 | Q 493 | E 35 | 0.694 |
| E 406 | D 30 | 0.76 | N 501 | Y 41 | 1 | Q 498 | K 353 | 0.99 | Q 493 | K 31 | 0.784 |
| R 403 | R 393 | 0.92 | R 403 | E 37 | 1.405 | E 406 | E 37 | 1.01 | R 403 | E 37 | 1.569 |
| E 406 | E 37 | 0.98 | T 500 | D 53 | 1.511 | D 405 | E 37 | 1.2 | K 458 | E 23 | 1.747 |
| D 405 | E 37 | 1.22 | K 458 | E 23 | 1.892 | T 500 | Y 41 | 1.27 | T 500 | D 355 | 2.738 |
| R 403 | K 353 | 2.3 | K 417 | D 30 | 4.944 | R 403 | K 353 | 2.39 | K 417 | D 30 | 2.938 |